\documentclass[pre,twocolumn,showpacs,superscriptaddress,amsmath,amssymb]{revtex4-1}
\usepackage{times}
\usepackage{graphicx}
\usepackage{dcolumn}
\usepackage{subfigure}
\usepackage{color}
\usepackage[dvipdfm,colorlinks,citecolor=blue,linkcolor=blue]{hyperref}

\begin{document}

\title{Complex synchronous behavior in interneuronal networks with delayed inhibitory and fast electrical synapses}

\author{Daqing Guo}
\email{dqguo07@gmail.com}
\affiliation{Key Laboratory for NeuroInformation of Ministry of Education, School of Life Science and Technology, University of Electronic Science and Technology of China, Chengdu 610054, P. R. China}
\affiliation{Computational Neuroscience Unit, Okinawa Institute of Science and Technology Graduate University, Okinawa 904-0411, Japan}

\author{Qingyun Wang}
\email{nmqingyun@163.com}
\affiliation{Department of Dynamics and Control, Beihang University, Beijing 100191, P. R. China}

\author{Matja\v{z} Perc}
\email{matjaz.perc@uni-mb.si}
\affiliation{Faculty of Natural Sciences and Mathematics, University of Maribor, Koro\v{s}ka cesta 160, SI-2000 Maribor, Slovenia}

\begin{abstract}
Networks of fast-spiking interneurons are crucial for the generation of neural oscillations in the brain. Here we study the synchronous behavior of interneuronal networks that are coupled by delayed inhibitory and fast electrical synapses. We find that both coupling modes play a crucial role by the synchronization of the network. In addition, delayed inhibitory synapses affect the emerging oscillatory patterns. By increasing the inhibitory synaptic delay, we observe a transition from regular to mixed oscillatory patterns at a critical value. We also examine how the unreliability of inhibitory synapses influences the emergence of synchronization and the oscillatory patterns. We find that low levels of reliability tend to destroy synchronization, and moreover, that interneuronal networks with long inhibitory synaptic delays require a minimal level of reliability for the mixed oscillatory pattern to be maintained.
\end{abstract}


\maketitle

\section{Introduction} \label{sec:1}
Synchronization is an important phenomenon that occurs in many biological and physical systems \cite{r1}. In the brain, the rhythmic oscillations of concerted electrical activity, which are representative for the synchronous firing of neurons, have been observed in different regions, including the neocortex, hippocampus and thalamus \cite{r2,r3}. Since neural oscillations are associated with many high-level brain functions, the pertinent research has attracted considerable attention in the past decades \cite{r2, r3, r4}. It has been proposed that these oscillations not only carry information by themselves, but that they may also regulate the flow of information and assist by its storage and retrieval in neural circuits \cite{r5}.

In the brain, fast-spiking interneurons are mutually connected by both inhibitory chemical synapses as well as electrical synapses (gap junctions) \cite{r6}. The evidence is mounting that networks composed of fast-spiking interneurons could provide synchronization mechanisms by means of which important rhythmic activities, such as the gamma ($\gamma$: 25-100 Hz) rhythm and the mixed theta ($\theta$: 4-12Hz) and gamma rhythm  \cite{r7,r7a1,r8}, can be generated. Computational studies indicate that inhibitory and electrical synapses play an important role by the generation of these oscillations. For example, it has been shown that interneuronal networks with solely inhibitory synapses can produce gamma oscillations, but that adding gap junctions to the network can further increases their stability and coherence \cite{r7a1}. It has also been reported that interneuronal networks coupled by both fast and slow inhibitory synapses can produce the mixed theta and gamma rhythmic activity \cite{r8}. In addition, several theoretical studies have been performed to provide a deeper understanding of how the inhibitory and electrical synapses promote synchronization amongst coupled neurons \cite{r9}.

Information transmission delays, which are due to the finite propagation speeds and due to time lapses occurring by both dendritic and synaptic processing \cite{r10}, are also an inherent part of neuronal dynamics. In particular the transmission delays of chemical synapses are not to be neglected \cite{r10}. Physiological experiments have revealed that they can be up to several tenths of milliseconds in length \cite{r11}. On the other hand, the transmission delays introduced by electrical synapses are comparably short, usually not exceeding 0.05 milliseconds \cite{r11}, so they are often not taken explicitly into account. In terms of dynamical complexity, the existence of time delays makes a nonlinear system with a finite number of degrees of freedom become an infinite-dimensional one, which may enrich the dynamics \cite{r12}, enhance synchronization \cite{r13}, and facilitate spatiotemporal pattern formation \cite{r14}.

Although existing studies attest clearly to the fact that information transmission delays have a significant impact on the synchronization of interneuronal networks, to the best of our knowledge the focus has always been on considering only short inhibitory synaptic delays \cite{r15}. However, since as noted above, the delays of chemical synapses may be substantial, it is also of interest to consider long inhibitory synaptic delays. Given that previous studies have shown that different time delay lengths may have rather different effects on the synchronization of coupled nonlinear oscillators \cite{r13, r14}, we anticipate that the consideration of long inhibitory synaptic delays may lead to new and inspiring results.

To resolve this, we here study the synchronization in an interneuronal network that is coupled by both delayed inhibitory as well as fast electrical synapses, focusing specifically on the effects of inhibitory synaptic delays covering a wide window of values. Our simulations reveal that the delayed inhibition not only plays an important role in network synchronization, but that it can also lead to different oscillatory patterns. The comparatively fast gap-junctional coupling, on the other hand, contributes solely to the synchronization of the network but does not affect the emergence of oscillatory patterns. Most interestingly, we show that a sufficiently long inhibitory synaptic delay induces a rapid transition from the one-frequency to the two-frequency state, thus leading to the occurrence of a mixed oscillatory pattern. Moreover, we also show that the unreliability of inhibitory synapses has a significant impact on both the synchronization and the emergence of oscillatory patterns. Our findings thus add to the established relevance of time delays in neuronal networks and highlight the importance of synaptic mechanisms for the generation of synchronized neural oscillations.

The reminder of this paper is organized as follows. In Section~\ref{sec:2}, we present the mathematical model and introduce the synchronization measure. Main results are presented in Section~\ref{sec:3}, while in Section~\ref{sec:4} we summarize our work and briefly discuss potential biological implications of our findings.

\section{Mathematical model}\label{sec:2}
\begin{figure}[!t]
\includegraphics[width=7cm]{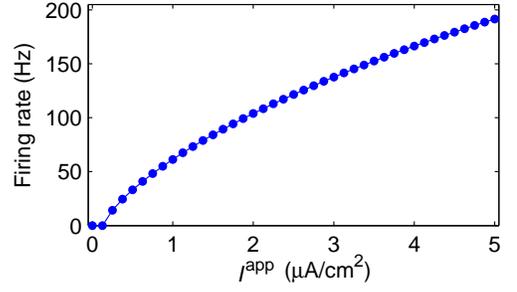}
\caption{\label{fig:1}(Color online) Firing rate curve of the Wang-Buzsaki model in dependence on the externally applied current $I^{\text{app}}$ in the absence of additional inputs.}
\end{figure}

We consider a network composed of $N$ fast-spiking interneurons. Neurons in the network are randomly connected by inhibitory and electrical synapses with probability $p_c$ and $p_e$, respectively. For simplicity, all synapses are bidirectional. We do not allow a neuron to be coupled to another neuron more than once by using the same type of synaptic coupling, or a neuron to be coupled with itself. We assume that all electrical synapses are fast, thus considering delays only by the inhibitory synapses. This assumption is reasonable, as we have argued in the Introduction. The dynamics of fast-spiking interneurons is described by the Wang-Buzsaki model \cite{r16}. It has a form similar to the classical Hodgkin-Huxley model \cite{r10,r17}, with details as follows \cite{r16}:
\begin{equation}
\begin{split}
C\frac{dV_i}{dt}=&-g_{\text{Na}}m_i^3h_i(V_i-E_{\text{Na}})-g_{\text{K}}n_i^4(V_i-E_{\text{K}})\\&-g_{\text{L}}(V_i-E_{\text{L}})+I_i^{\text{app}}+I_i^{\text{syn}}(t),
\end{split}
\label{eq:1}
\end{equation}
where the three gating variables obey the following equations
\begin{equation}
\begin{split}
m_i&=\alpha_{m_i}(V_i)/\left[\alpha_{m_i}(V_i)+\beta_{m_i}(V_i)\right],\\
\frac{dh_i}{dt}&=\phi\left[\alpha_{h_i}(V_i)(1-h_i)-\beta_{h_i}(V_i)h_i\right],\\
\frac{dn_i}{dt}&=\phi\left[\alpha_{n_i}(V_i)(1-n_i)-\beta_{n_i}(V_i)n_i\right].
\end{split}
\label{eq:2}
\end{equation}
Here $i=1, 2, ..., N$ is the neuron index, $V_i$ denotes the membrane potential of neuron $i$, and the six rate functions are \cite{r16}: $\alpha_{m_i}(V_i)=0.1(V_i+35)/(1-\exp{(-0.1(V_i+35))})$, $\beta_{m_i}(V_i)=4\exp{(-(V_i+60)/18)}$, $\alpha_{h_i}(V_i)=0.07\exp{(-(V_i+58)/20)}$, $\beta_{h_i}(V_i)=1/(\exp{(-0.1(V_i+28))}+1)$, $\alpha_{n_i}(V_i)=0.01(V_i+34)/(1-\exp{(-0.1(V_i+34))})$, and $\beta_{n_i}(V_i)=0.125\exp{(-(V_i+44)/80)}$. $I_i^{\text{syn}}$ is the synaptic current of neuron $i$ due to the interactions with other neurons within the network (also referred to as internal synaptic current in this paper), and $I_i^{\text{app}}$ is an externally applied current representing the collective effect of inputs coming from the outside of the network. In this work, we model the externally applied current as $I_i^{\text{app}}=I_0+\sigma\mu_i(t)$, where $I_0$ is the mean current, $\mu_i(t)$ is an independent Gaussian white noise with zero mean and unit variance, and $\sigma$ is the intensity of stochastic fluctuations. The parameters of the Wang-Buzsaki model assume standard values \cite{r16}: $C=1$ $\mu$F/cm$^2$, $g_{\text{Na}}=35$ ms/cm$^2$, $E_{\text{Na}}=55$ mV, $g_{\text{K}}=9$ ms/cm$^2$, $E_{\text{K}}=-90$ mV, $g_{\text{L}}=0.1$ ms/cm$^2$, $E_{\text{L}}=-65$ mV, and $\phi=5$. A spike is detected whenever the membrane potential exceeds the threshold of $V_{\text{th}}=-10$ mV. Figure~\ref{fig:1} shows the firing rate curve of the Wang-Buzsaki model when the later is driven solely by the externally applied current ($\sigma=0$).

For each neuron, the internal synaptic current consists of two terms, which are
\begin{equation}
\begin{split}
I_i^{\text{syn}}(t)= \sum\nolimits_{j}w_{ij} r_{ij} [E_{\text{inh}}-V_i] + \sum\nolimits_{k}g_{ik}[V_k-V_i].
 \end{split}
 \label{eq:3}
\end{equation}
In this equation, the first and second outer sums run over all inhibitory and electrical synapses onto neuron $i$, $w_{ij}$ is the inhibitory synaptic strength from neuron $j$ to neuron $i$, $r_{ij}$ is the corresponding inhibitory synaptic variable,  $E_{\text{inh}}=-80$ mV is the reversal potential for inhibitory synapses, and $g_{ik}$ is the electrical synaptic strength from neuron $k$ to neuron $i$. For inhibitory synapses, once a presynaptic neuron emits a spike, the corresponding $r_{ij}$ is updated after a fixed spike transmission delay $\tau_{ij}$, according to $r_{ij} \leftarrow r_{ij}+1$. Otherwise $r_{ij}$ decays exponentially with a fixed time constant $\tau_s$. For simplicity, we set $w_{ij}=w$, $\tau_{ij}=\tau$ and $g_{ik}=g$ throughout this paper, implying that the coupling is identical for the same type of synapses.

To characterize the synchronization within the network, a dimensionless synchronization measure $S$ is introduced, following \cite{r18}. We first compute the time fluctuations of the average membrane potential according to
\begin{equation}
\begin{split}
\Delta_N=\langle A_N(t)^2\rangle _t - \langle A_N(t)\rangle _t^2,
 \end{split}
 \label{eq:4}
\end{equation}
where the sign $\langle\cdot\rangle_t$ denotes the average over time and $A_N(t)=\sum_{i=1}^NV_i(t)/N$ is the average membrane potential at time $t$. Subsequently, the population-averaged variance of the activity of each individual neuron is determined according to
\begin{equation}
\begin{split}
\Delta=\frac{1}{N}\sum_{i=1}^N\left(\langle V_i(t)^2\rangle _t-\langle V_i(t)\rangle _t^2\right).
 \end{split}
 \label{eq:5}
\end{equation}
Finally, the synchronization measure is computed as
\begin{equation}
\begin{split}
S=\frac{\Delta_N}{\Delta}.
 \end{split}
 \label{eq:6}
\end{equation}
From this it follows that the larger the value of $S$ the better the synchronization in the network.

The described mathematical model is integrated numerically using the fourth-order Runge-Kutta algorithm with a fixed time step of $h=0.025$ ms. This is sufficiently small to ensure an accurate simulation of the Wang-Buzsaki model \cite{r16}. For each set of parameters, the initial membrane potentials of neurons are uniformly distributed between -70 and 30 mV. The network size is $N=300$, while $p_c=0.1$ and $p_e=0.05$. We always generate both types of synapses, but use $g=0$ to denote the network without the fast gap-junctional coupling. The two parameters that determine the externally applied current are $I_0=1.4$ $\mu$A/cm$^2$ and $\sigma=0.25$ $\mu$A ms$^{1/2}$/cm$^2$. Under these conditions, the mean firing rate of the Wang-Buzsaki model is approximately 80 Hz in the absence of the internal synaptic current. We perform all simulations up to 3000 ms, and collect the data from 1000 to 3000 ms for further statistical analysis. The reported results, except for the spike raster diagrams, are averages over 30 independent runs.

\section{Results} \label{sec:3}
\begin{figure}[!t]
\includegraphics[width=8.8cm]{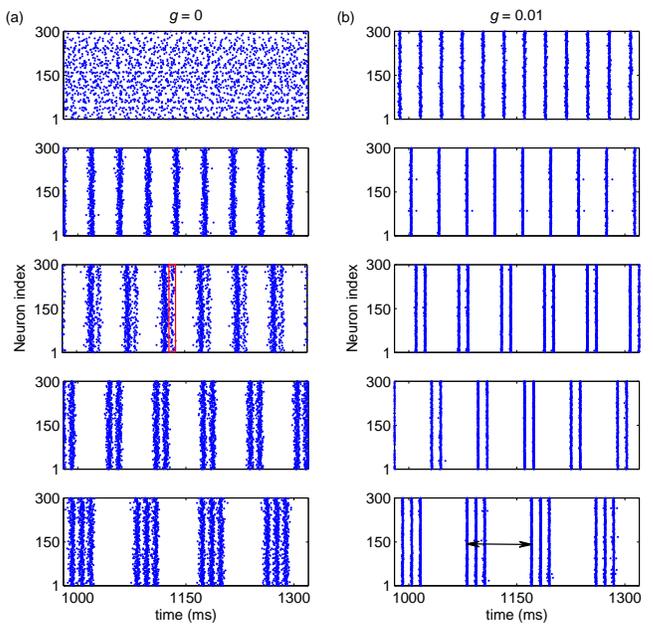}
\caption{\label{fig:2}(Color online) Spike raster diagrams for different values of the inhibitory synaptic delay $\tau$. (a) Without fast electrical synapses ($g=0$). (b) With fast electrical synapses ($g=0.01$ ms/cm$^2$). In all cases we set $w=0.01$ ms/cm$^2$ and $\tau_s=10$ ms. From top to bottom, $\tau$ = 0, 7, 13, 18 and 30 ms, respectively. The red rectangle in the middle panel of (a) illustrates that, without the fast gap-junctional coupling, the synchronous firing of the last group in each periodic cycle is weak when $\tau$ is near the transition point. The black double arrow line in the bottom panel of (b) denotes one periodic cycle for the mixed oscillatory pattern.}
\end{figure}

\begin{figure}[!t]
\includegraphics[width=8cm]{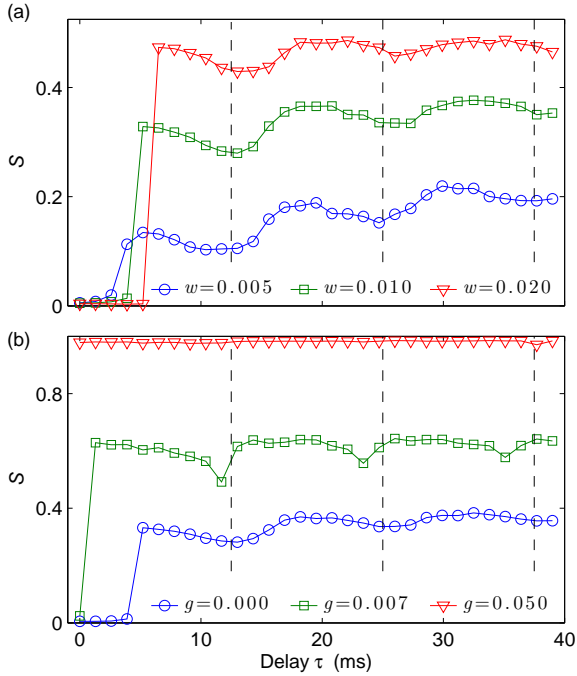}
\caption{\label{fig:3}(Color online) (a) Dependence of the synchronization measure $S$ on the delay $\tau$ for different values of the inhibitory synaptic strength $w$. Here $g=0$. (b) Dependence of $S$ on $\tau$ for different values of the electrical synaptic strength $g$. Here $w=0.01$ ms/cm$^2$. In all cases we use $\tau_s=10$ ms. The units of parameters $w$ and $g$ shown in panels (a) and (b) are ms/cm$^2$. The positions of the vertical dashed lines are: $\tau=12.5$, 25 and 37.5 ms, respectively.}
\end{figure}

We first show elementary simulation results that reveal how the delayed inhibitory and fast gap-junctional coupling influence the synchronous behavior of the considered interneuronal network. In Figs.~\ref{fig:2}(a) and \ref{fig:2}(b), several typical spike raster diagrams for different values of the inhibitory synaptic delay $\tau$, without ($g=0$) and with ($g\neq0$) the fast electrical synapses, are plotted, respectively. Presented results show clearly that delayed inhibitory as well as fast electrical synapses play an important role by the synchronization of the network. Without the fast gap-junctional coupling, the neuronal firings at $\tau=0$ are rather disordered. Essentially this is because the considered network ($p_c=0.1$) is sparse, i.e., there are not much more links between the neurons (on average) than there are neurons constituting the network, and hence cannot be easily synchronized in the absence of additional mechanisms that promote the onset of synchronization. By introducing inhibitory synaptic delays, an improvement in the synchronization of the network can be observed. However, it can also be observed that this depends significantly on the length of the inhibitory synaptic delay. Only suitable delays can help the network to maintain a high level of synchronization (compare results obtained with $\tau=7$, 18 and 30 ms, and $\tau=13$ ms in Fig.~\ref{fig:2}(a)). We demonstrate this quantitatively in Fig.~\ref{fig:3}(a), where also a near periodic oscillatory behavior in $S$ can be detected, which may be related to the matching of the inherent neuronal time scales with the duration of the delay, as proposed in \cite{r13} (we will discuss this near-periodic oscillatory behavior further shortly). Furthermore, the results presented in Fig.~\ref{fig:2}(b) suggest that the fast electrical synapses provide a strong mechanism for fostering synchronization. With the fast gap-junctional coupling turned on, the high-quality synchronization may be observed even at $\tau=0$. As the strength of the electrical synaptic coupling ($g$) is increased, the neuronal firings become more and more synchronized (see Fig.~\ref{fig:3}(b)). For sufficiently strong $g$, the measure $S$ approaches 1, indicating that the synchronization is almost perfect. Indeed, several previous studies have concluded that gap-junctional coupling is more effective than chemical coupling in leading to highly synchronized states \cite{r19}. One possible mechanism for this is that chemical synapses only act while the presynaptic neuron is spiking, whereas the electrical synapses are more efficient and can transmit the membrane potentials of presynaptic neurons to the corresponding postsynaptic neurons at all times.

\begin{figure}[!t]
\includegraphics[width=8cm]{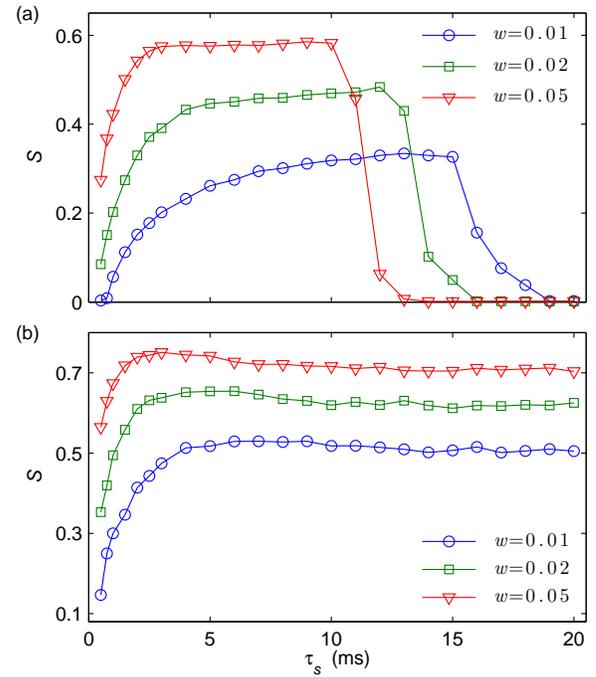}
\caption{\label{fig:4}(Color online) Dependence of the synchronization measure $S$ on the inhibitory synaptic time constant $\tau_s$ for different values of the inhibitory synaptic strength. (a) Without the fast electrical synapses ($g=0$), and (b) with the fast electrical synapses ($g=0.005$ ms/cm$^2$). In all cases we set $\tau=8$ ms. The values of the inhibitory synaptic strength considered here are $w=0.01$, 0.02, and 0.05 ms/cm$^2$, respectively.}
\end{figure}

Results presented in Fig.~\ref{fig:2} reveal also that the oscillatory pattern is largely influenced by the inhibitory synaptic delay. For suitably short values of $\tau$ a regular oscillatory pattern can be observed. Interestingly however, if the delay $\tau$ is sufficiently long, we can observe the emergence of a mixed oscillatory pattern, which implies that there are two main oscillation frequencies present in the network. One is the low frequency of the whole mixed oscillatory pattern and the other is the high frequency of the fast oscillations within each periodic cycle (see Fig.~\ref{fig:2}). Theoretically, the mixed oscillatory pattern appears only for sufficiently long inhibitory synaptic delay. Under this condition, neurons in the synchronized or near-synchronized network have enough time to fire more than once during a whole periodic cycle, before the inhibitory synaptic currents caused by the first synchronous spiking group within the same periodic cycle start to suppress their firing. Obviously, the longer the inhibitory synaptic delay $\tau$, the more groups of synchronous spikes might be contained in each periodic cycle (see $\tau=18$ and 30 ms in Figs.~\ref{fig:2}(a) and \ref{fig:2}(b)). Moreover, once the inhibitory synaptic currents caused by the first synchronous spiking group start to have effect, these currents tend to decrease the membrane potentials of neurons and prolong their firing period. In this case, the following inhibitory synaptic bombardments due to one or more synchronous spiking groups from the same periodic cycle will further suppress neuronal firings, and therefore the neurons in the network can fire again only after these inhibitory effects wear down or become fully absent. This provides a viable mechanism for the emergence of the low-frequency component in the mixed oscillatory pattern. While performing additional simulations, we have discovered that without the long delayed inhibitory synapses, networks of interneurons cannot generate the mixed oscillatory pattern even if we consider the non-physiological case of long electrical synaptic delays (data not shown). The above results thus indicate that the inhibitory synaptic delay serves as an important control parameter for the selection of the oscillatory pattern in the network, and that long inhibitory synaptic delays provide a stable mechanism for the emergence of the mixed oscillatory pattern in interneuronal networks.

We now return to Fig.~\ref{fig:3}(a) and further discuss the near-periodic oscillatory  behavior in $S$. It can be determined that the frequency of this near-periodic behavior matches with the oscillation frequency of the high-frequency component quite well. Note that we will show later that the oscillation frequency of the high-frequency component is mainly influenced by the parameter $\tau_s$. In the narrow regions of $\tau$ where the minima of $S$ appear (note that these regions also correspond to the transition points for the number of synchronously spiking groups contained in each periodic cycle), only a limited amount of neurons will participate in the last group of synchronous firings within each periodic cycle (see the red rectangle in the middle panel of Fig.~\ref{fig:3}(a)). This can be attributed to the matching between the synaptic delay $\tau$ and the high-frequency component, which ultimately causes the near-periodic behavior in the synchronization measure $S$. We have also performed additional numerical simulations by using other models of neuronal dynamics, such as the model by Izhikevich with fast-spiking dynamics \cite{r20} and the standard Hodgkin-Huxley model \cite{r10,r17}, and we have observed qualitatively identical results, thus confirming the generality of this phenomenon (see also \cite{r13}). On the other hand, our results also reveal that the fast gap-junctional coupling tends to suppress the occurrence of such near-periodic oscillations, which may be attributed to the overall promotion of synchronization (see Figs.~\ref{fig:3}(b)). Indeed, if the electrical coupling is sufficiently strong this phenomenon disappears altogether because then the neurons in the network are perfectly synchronized.

In addition to the time delay $\tau$, we also find that the synchronization depends significantly on the other two important inhibitory synaptic parameters, which are the strength $w$ and the time constant $\tau_s$. Figures~\ref{fig:4}(a) and \ref{fig:4}(b) depict the synchronization measure $S$ as a function of $\tau_s$ for different values of $w$, without ($g=0$) and with ($g\neq0$) the fast electrical synapses, respectively. In the absence of fast gap-junctional coupling, there exists an optimal region of $\tau_s$ in each depicted dependence of $S$, which implies that the network can support synchronization optimally only for intermediate values of $\tau_s$ (see also \cite{r17} for related results). In this case, a strong inhibitory synaptic strength can drive the network towards a high-level of synchronization at the corresponding optimal value of $\tau_s$. However, with the increasing of $w$, it can also be observed that the top plateau region of the $S$ curve becomes narrower and shifts to the left (the direction of short $\tau_s$). An explanation why longer $\tau_s$ can no longer produce high values of $S$ is as follows. Due to the heterogeneity of connectivity, some neurons in the considered network will have more inhibitory synaptic inputs than others. For long $\tau_s$, the slowly decaying synaptic inhibition accumulates in time and thus may lead to a tonic level of hyperpolarizing currents that cancel the external depolarizing currents by neurons with more inhibitory inputs \cite{r17}. This will suppress or even fully disable the firing of such neurons, which in turn means that if the synaptic time constant is too long, the synchronization will deteriorate significantly. At a fixed $\tau_s$, a large value of $w$ will introduce more inhibition to the network, and thus a relatively shorter $\tau_s$ will be needed to impair synchronization. As a result, although strong $w$ can enhance the synchronization in the corresponding optimal $\tau_s$ region, they may also reduce the size of this region. On the other hand, adding the fast gap-junctional coupling to the network, as expected, will enhance the synchronous firing of neurons. This enhancement is quite remarkable, even if the gap-junctional coupling is rather weak, as shown in Fig.~\ref{fig:4}(b). Further increasing the strength of gap-junctional coupling can lead to the perfect synchronization (data not shown). Several previous experiments have shown that some of the inhibitory synapses between interneurons can have rather slow synaptic kinetics \cite{r8}. Therefore, to some extent, our results suggest that the fast electrical synapses might be essential for taming desynchronization if an interneuronal network contains a considerable number of slow inhibitory synapses.

\begin{figure}[!t]
\includegraphics[width=8.8cm]{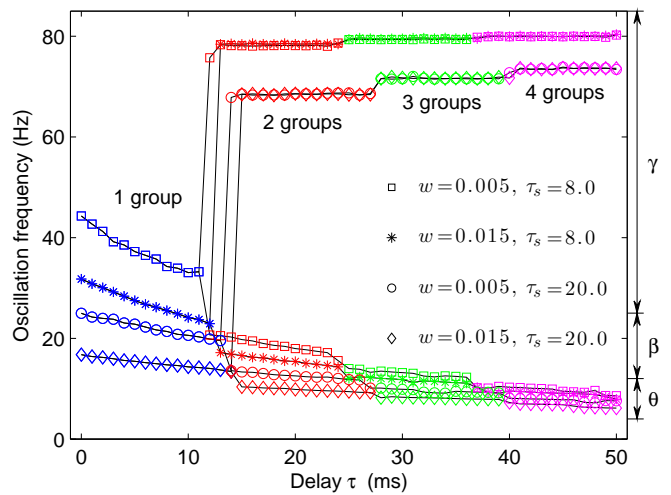}
\caption{\label{fig:5}(Color online) Dependence of the oscillation frequency on the delay $\tau$ for different levels of inhibition. In all cases we set $g=0.03$ ms/cm$^2$. The units of parameters $w$ and $\tau_s$ shown here are ms/cm$^2$ and ms, respectively. The oscillation frequency is divided into three bands: theta band ($\theta$: 4-12 Hz), beta band ($\beta$: 12-25 Hz), and gamma band ($\gamma$: 25-100 Hz). Different colors (shades of gray) refer to different numbers of synchronously spiking groups contained in each periodic cycle.}
\end{figure}

Figure~\ref{fig:5} shows how the oscillation frequency depends on the inhibitory synaptic delay for different levels of inhibition. As can be observed, the oscillation frequency of the considered interneuronal network is determined by both the inhibitory synaptic delay as well as the inhibition. In the short $\tau$ region, the oscillations are characterized by a single frequency. In this case, increasing $w$ and $\tau_s$ can reduce the oscillation frequency from the $\gamma$ band to $\theta$ band (see Fig.~\ref{fig:5}). Once $\tau$ exceeds a critical time delay $\tau_c$, we observe that the network oscillations transit from the one-frequency state to the two-frequency state, indicating the emergence of the mixed oscillatory pattern. This transition is rapid and stable with the aid of fast gap-junctional coupling. For the mixed oscillatory pattern, our results show that both $w$ and $\tau_s$ influence the low-frequency component, but only $\tau_s$ has a significant effect on the high-frequency component. When $\tau_s$ is short, the inhibitory synaptic currents from one periodic cycle decay fast, so that they may completely vanish before the firing of neurons enters into the next periodic cycle and almost do not influence the neuronal firing in the next periodic cycle. Thus, in this case, the critical time delay $\tau_c$ is approximately 12.5 ms, and based on the same reason, the number of synchronous spiking groups contained in each periodic cycle is also increased once about every 12.5 ms. The above analysis suggests that the oscillation frequency of the high-frequency component is around 80 Hz (in the $\gamma$ band) for short $\tau_s$ (see $\tau_s=8$ ms in Fig~\ref{fig:5}), corresponding to the firing rate of a single Wang-Buzsaki neuron that is driven solely by the considered externally applied current. If $\tau_s$ is sufficiently long, the inhibitory synaptic currents from one periodic cycle can persist to a certain extent even after the neuronal firing enters into the next periodic cycle. These remaining inhibitory synaptic currents will suppress the neuronal firing in the next periodic cycle, and thus increase the firing interval between the first and the second synchronous spiking groups in the next periodic cycle. Therefore, the firing intervals of the high-frequency component are not perfectly identical, i.e., the first firing interval is slighter larger than the other firing intervals. This in turn yields a relatively smaller average frequency of the high-frequency component. As a result, for long $\tau_s$ the system needs a relatively longer $\tau$ to generate the mixed oscillatory pattern, and it also exhibits a slightly smaller frequency in the high-frequency component of its output (see $\tau_s=20$ ms in Fig~\ref{fig:5}). Moreover, our results also show that the frequency of the whole mixed oscillatory pattern is quite low, even in the case of weak inhibition (small $g$ and short values of $\tau_s$). For sufficiently long values of $\tau$ this oscillation frequency can be maintained in the $\theta$ band quite efficiently. The mixed theta and gamma rhythm is believed to play an important role in brain cognitive functions \cite{r21}. The traditional viewpoint is that interneuronal networks with fast and slow inhibitory synaptic dynamics are the basic neural circuits to generate this special type of neural oscillations \cite{r8}. Our results provide a new insight related to this, which is that long transmission delays of inhibitory synapses may also lead to the mixed theta and gamma rhythm in interneuronal networks. We note that this mechanism is still functional even if only some (not all) of the inhibitory synapses are considered to have long transmission delays, provided only that the delayed inhibitory synaptic currents are strong enough.

\begin{figure}[!t]
\includegraphics[width=8.8cm]{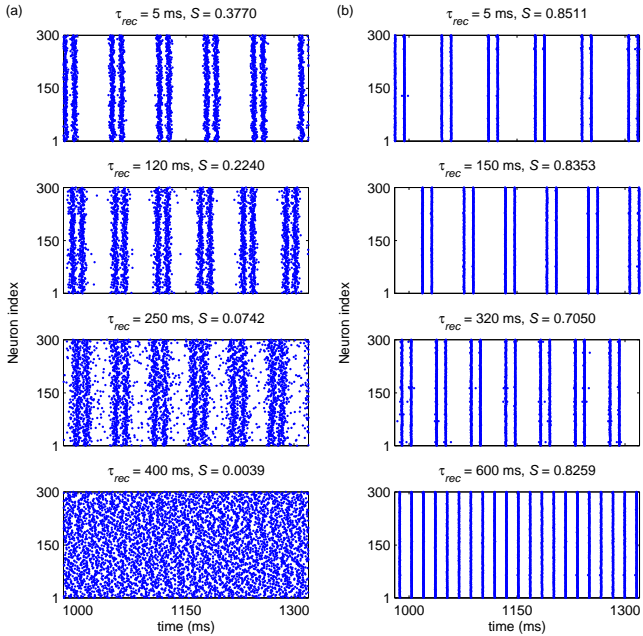}
\caption{\label{fig:6}(Color online) Effects of the time constant of the recovered variable $\tau_{rec}$ on the synchronization and the emergence of oscillatory pattern in the interneuronal network. (a) Without the fast electrical synapses ($g=0$). (b) With the fast electrical synapses ($g=0.02$ ms/cm$^2$). In all cases we set $\tau=18$ ms, $w=0.05$ ms/cm$^2$, $\tau_s=10$ ms, $\tau_{in}=3$ ms, and $u_0=0.2$. Left panel: from top to bottom, $\tau_{rec}=5$, 120, 250, and 400 ms, respectively. Right panel: from top to bottom, $\tau_{rec}=5$, 150, 320, and 600 ms, respectively.}
\end{figure}

Finally, we examine how the unreliability of inhibitory synapses influences the synchronization and oscillatory patterns in the studied interneuronal network. This investigation is carried out because the synaptic transmission through real chemical synapses is indeed to a degree unreliable \cite{r22}, and also because several previous studies have advocated that the unreliable synapses may play important functional roles in neural computation \cite{r23}. In principle, the unreliability of chemical synapses can be explained by the phenomenon of probabilistic transmitter release, which has been confirmed by biological experiments \cite{r24}. Typically, the synaptic unreliability is associated with synaptic depression, which can be simulated by a well-established phenomenological model proposed in \cite{r25}. In this model, three parameters $x_{ij}$, $y_{ij}$, and $z_{ij}$, which denote the fractions of synaptic resources in the recovered, active, and inactive states, are employed and their dynamical equations are given by: $\dot{x}_{ij}=z_{ij}/\tau_{rec}-u_0 \cdot x_{ij}  \cdot \delta(t-t_j^k-\tau)$, $\dot{y}_{ij}=-{y_{ij}}/{\tau_{in}}+u_0  \cdot x_{ij}  \cdot \delta(t-t_j^k-\tau)$, and $\dot{z}_{ij}={y_{ij}}/{\tau_{in}}-{z_{ij}}/{\tau_{rec}}$. Here $\delta(t)$ is the Dirac delta function, $t_j^k$ gives the timing of presynaptic spikes, $\tau_{in}$ is the time constant of the inactive variable, $\tau_{rec}$ is the time constant of the recovered variable, and $u_0$ describes the utilization of synaptic efficacy. Here we apply this model to modulate the updating of synaptic conductance as follows: whenever a presynaptic neuron $j$ fires a spike, the corresponding postsynaptic conductances are increased instantaneously after a fixed spike transmission delay $\tau$, according to $r_{ij} \leftarrow r_{ij}+y_{ij}(t)$; otherwise $r_{ij}$ decays exponentially with a fixed time constant $\tau_s$. In the following simulations, we set $\tau_{in}=3$ ms and $u_0=0.2$, and change the variable $\tau_{rec}$ to control the synaptic depression. A longer $\tau_{rec}$ corresponds to a stronger synaptic depression, and therefore denotes a lower level of synaptic reliability.

\begin{figure}[!t]
\includegraphics[width=8.3cm]{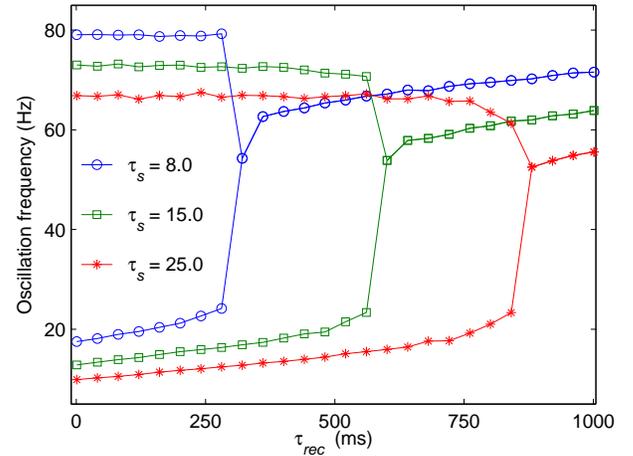}
\caption{\label{fig:7}(Color online) Dependence of the oscillation frequency on $\tau_{rec}$ for different values of $\tau_{s}$. In all cases we set $g=0.02$ ms/cm$^2$, $w=0.05$ ms/cm$^2$, $\tau=15$ ms, $\tau_{in}=3$ ms, and $u_0=0.2$. The values of $\tau_{s}$ considered here are $\tau_{s}=8$, 15, and 25 ms, respectively.}
\end{figure}

In Figs.~\ref{fig:6}(a) and \ref{fig:6}(b), we present several typical spike raster diagrams for different values of  $\tau_{rec}$, without ($g=0$) and with ($g\neq0$) the fast electrical synapses, respectively. We choose the inhibitory synaptic delay to equal $\tau=18$ ms, ensuring that the network will be in the mixed oscillatory pattern when $\tau_{rec} \to 0$ ms. Results presented in Fig.~\ref{fig:6} demonstrate that the unreliability of inhibitory synapses has a great impact on both the network synchronization and the emergence of oscillatory patterns. In the absence of fast gap-junctional coupling, the synchronization reduces markedly with increasing $\tau_{rec}$. In this case, high synaptic unreliability (long $\tau_{rec}$) leads to insufficient synaptic information interaction, which largely deteriorates network synchronization and causes the neural oscillations to disappear completely (see $\tau_{rec}=400$ ms in Fig.~\ref{fig:6}(a)). With the fast gap-junctional coupling incorporated, we find that the network synchronization can be maintained in the majority of the $\tau_{rec}$ region. Again, this is because the gap-junctional coupling itself can provide an effective mechanism for network synchronization. However, our results also show that the considered network needs a certain level of inhibitory synaptic reliability for the mixed oscillatory pattern to be preserved. For sufficiently long $\tau_{rec}$, it can be observed that the mixed oscillatory pattern transforms to the regular oscillatory pattern due to the lack of inhibition (see $\tau_{rec}=600$ ms in Fig.~\ref{fig:6}(b)). The transitions in the oscillatory patterns can be observed more clearly from the data presented in Fig.~\ref{fig:7}. These findings suggest that the unreliability of inhibitory synapses might also provide a flexible mechanism for controlling the switch between different oscillatory patterns in interneuronal networks.

\section{Discussion} \label{sec:4}
In summary, we have employed a computational approach with the aim of investigating the complex synchronous behavior in interneuronal networks that are coupled by delayed inhibitory and fast electrical synapses. We have shown that these two types of synaptic coupling play an important role in warranting network synchronization. In particular, the considered network can achieve a high level of synchronization either by means of a suitable tuning of the inhibitory synaptic delay, by enhancing the strength of electrical synapses, or by means of both. On the other hand, our simulations have revealed that only delayed inhibition significantly influences the emergence of oscillatory patterns, while electrical synapses play at most a side role by this phenomenon. In particular, we have shown that short inhibitory delays evoke regular oscillatory patterns, while sufficiently long delays can lead to an abrupt emergence of mixed oscillatory pattern. By analyzing the oscillation frequencies, we found that the considered interneuronal network can generate both types of oscillations in physiologically relevant frequency bands, such as the gamma rhythm and the mixed theta and gamma rhythm. This fact might have biological implications as these rhythmic activities are frequently associated with fast-spiking interneurons, and are also believed to play prominent functional roles in cognitive tasks \cite{r7,r8,r21}. Lastly, we have also demonstrated that the unreliability of inhibitory synapses plays an important role by the synchronization of the network as well as by the emergence of oscillatory patterns. More precisely, we have shown that high levels of unreliability destroy synchronization, and that a minimal level of reliability is needed for the emergence and stability of the mixed oscillatory pattern.

We hope that the presented results will improve our understanding of the synaptic mechanisms that are responsible for the generation of synchronous oscillations in the neural tissue. Indeed, our findings suggest that delayed inhibitory synapses are a viable candidate for controlling the emergence of oscillatory patterns. Depending on the actual biological circumstances, the same interneuronal ensembles may produce neural oscillations with different patterns in an adaptive way through the modulation of synaptic transmission. We also hope that this study will inspire further research on this topic, in particular by taking into account additional physiological properties of neuronal networks, such as the anatomical connectivity and distance-dependent synaptic information transmission delays.

\begin{acknowledgments}
This research was supported by the National Natural Science Foundation of China (Grant No. 11172017) and the Slovenian Research Agency (Grant No. J1-4055). D. G. acknowledges the financial support from the University of Electronic Science and Technology of China.
\end{acknowledgments}

\end{document}